\documentclass[%
 reprint,
superscriptaddress,
 amsmath,amssymb,
 aps,
]{revtex4-2}

\usepackage{graphicx}
\usepackage{epstopdf}
\usepackage{dcolumn}
\usepackage{bm}
\newcommand\Rm{{\text{Rm}}}

\usepackage{caption}
\usepackage{subcaption}

\usepackage{xcolor}
\usepackage{hyperref}
\hypersetup{colorlinks=true}
\newcommand{\sv}[1]{\textcolor{black}{#1}}

\begin{document}

\preprint{APS/123-QED}

\title{Effects of radial conductivity variation on the Ponomarenko dynamo}

\author{Shipra Verma}
\email{shipraverma1997@kgpian.iitkgp.ac.in}
\affiliation{Department of Physics, Indian Institute of Technology Kharagpur, West Bengal, 721302, India}
\author{Kannabiran Seshasayanan}
\affiliation{Department of Applied Mechanics $\&$ Biomedical Engineering, Indian Institute of Technology Madras, Chennai, 600036, India}

\begin{abstract}

We study the effects of conductivity variation on the Ponomarenko dynamo. Taking monotonically increasing/decreasing and sinusoidally varying radial profiles, we study the kinematic dynamo problem. The threshold of the dynamo, given by the critical magnetic Reynolds number Rm$_c$, is found to strongly depend on the form of conductivity variation near the discontinuity of the velocity field where the shear is dominant. For monotonically varying profiles in the limit of sharp variation, this threshold maps to the problem of the dynamo with a jump in conductivity, while for the sinusoidal profile with large variations/wavenumbers, the threshold increases due to effective diffusivity arising from the rapid changes in conductivity. Far from the threshold, it is found that the growth rate depends only on the local gradient of the conductivity near the discontinuity, with the \sv{opposite} sign of shear and conductivity gradient leading to a larger growth rate, thereby aiding the dynamo instability in regenerating B$_{\phi}$ from B$_r$. We calculate the expression for the growth rate in the limit of large $\Rm \mbox{ }(\gg 1)$ and find that the conductivity variation leads to a modification proportional to the local gradient of conductivity times $\Rm^{-1/2}$. This correction is found to depend only on the local gradient of the conductivity at the discontinuity. Similar dependencies are found for realistic, smooth velocity profiles and for different magnetic boundary conditions.%
\end{abstract}

\maketitle


\section{\label{sec:level1} Introduction
}

The spontaneous growth of the magnetic field generated by the flow of conducting fluid, known as the dynamo effect, explains the existence and sustenance of the astrophysical magnetic field. In the dynamo problem, the field grows exponentially beyond a threshold, typically quantified using the magnetic Reynolds number $\Rm$. The growth rate and the saturation of the field depend on the underlying physical parameters, such as the magnetic and thermal diffusivities and viscosity of the conducting fluid, among many others. \sv{While early studies of the dynamo effect treated diffusivities as constants, the interiors of many planetary and stellar objects exhibit strongly varying viscosity and conductivity. These variations primarily arise due to the large, predominantly radial temperature/compositional gradients seen in such objects \cite{wicht2019dynamo,gomez2010effects,dietrich2018anelastic}. Therefore, to accurately model astrophysical systems, numerical dynamo models incorporate radially dependent viscosity and conductivity profiles, which have been found to affect both the hydrodynamic flow and the electromagnetic field \cite{duarte2018physical, dietrich2018anelastic, duarte2013anelastic, soyuer2021linking, gomez2010effects}. However, due to their inherently turbulent nature, large-scale mixing may homogenize the radial profiles. Using mixing length theory, commonly applied in stellar convection modelling, one can argue that such mixing would not significantly alter the background conductivity \cite{biermann1932untersuchungen, Pinsonneault_Ryden_2023}.} Typical \sv{radial} conductivity profiles that have been used in the models of planetary/stellar interiors consider monotonically increasing or decreasing profiles, which include regions of sharp conductivity variations \cite{duarte2018physical, wicht2019dynamo}. Other models have used the spatial periodic structure of the conductivity profile to study the dynamo effect \cite{rogers2017hottest}. It has been shown that variations in conductivity give rise to new mechanisms for the dynamo \cite{petrelis2016fluctuations}. Rapid variations in a short spatial region result in the localisation of the magnetic field and also affect the zonal flows in planetary models \cite{duarte2018physical, wicht2019dynamo, wulff2024effects}. Also, enhanced growth rates are seen in flows in Taylor-Couette geometries, \cite{marcotte2021enhanced}. 

Recent studies focus on the effects of conductivity variations in the geodynamo models using large-scale simulations \cite{gubbins2015core, dietrich2018anelastic, wicht2019dynamo}. While these models capture the dynamics in the relevant geometry, the high computational cost of such simulations limits our ability to systematically study the effects of the radial conductivity variation on the dynamo. Similarly, if one is interested in studying their effect at large $\Rm$ values, numerical simulations are constrained to work in the moderate range of parameters. Theoretical dynamo models, on the other hand, can allow one to study the dynamo problem in a more controlled setting with the help of both analytical and numerical methods. One of the earliest examples of such a theoretical model was proposed by Ponomarenko \cite{ponomarenko1973theory}, where a helical flow in a cylindrical geometry was studied. The experimental study \cite{gailitis2000detection} motivated by this theoretical flow successfully produced a self-amplified magnetic field. The Ponomarenko dynamo with the discontinuous velocity profile has been shown to be a fast dynamo \cite{gilbert1988fast} with the growth rate saturating to a non-zero value in the infinite $\Rm$ limit. This occurs due to the infinite shear at the discontinuity, which allows for the magnetic field to be concentrated at a very small length scale. For smooth velocity profiles, it is known that the dynamo growth tends to zero in the limit of large $\Rm$, thus an example of a slow dynamo.  The effect of magnetic boundary conditions in the Ponomarenko dynamo has been studied in detail \cite{kaiser1999kinematic, avalos2003influence}, and it is found that the threshold/growth rate of the dynamo strongly depends on it.

In this study, we quantify the modifications in the threshold and the growth rate of the Ponomarenko dynamo due to the radial variations in conductivity. \sv{We focus on the kinematic phase where the Lorentz force is negligible and does not alter the background velocity and, in turn, the conductivity. Additionally, we attribute conductivity variations to result solely from the background temperature variations}. Motivated by the earlier works, radial conductivity profiles, which are monotonically increasing/decreasing functions and sinusoidal functions of the radius, are considered. These profiles are defined with a free parameter that controls the steepness of the variation. With this setup, the quantification is done for a wide range of $\Rm$ values and for different steepness of the conductivity profiles. The detailed description of the problem and methodology used are discussed in Section \ref{sec:setup}. The results from numerical and theoretical approaches close and far from the onset of the dynamo instability are presented in Sections \ref{sec:onset} and \ref{sec:largeRm}. The effect of radial conductivity variation in the presence of a smooth velocity profile and for different magnetic boundary conditions is discussed in Section \ref{sec:mbc}. Concluding remarks are presented in Section \ref{sec:Conclusion}.

\section{Setup} \label{sec:setup}

We consider the Ponomarenko flow given by the velocity profile, 
\begin{align}
    {\bf u} = \left\{
    \begin{array}{cc}
         r \Omega (r) {\bf e}_\phi + U_z (r) {\bf e}_z & \mbox{if } r \leq r_0 \\
         0 & \mbox{if } r > r_0 
    \end{array},
    \right.
\end{align}
which represents a helical flow inside a cylinder of radius $r_0$. The induction equation gives the evolution of the magnetic field,
\begin{align}
    \partial_t {\bf B} = {\boldsymbol \nabla} \times \left( {\bf u} \times {\bf B} \right) - {\boldsymbol \nabla} \times \left( \eta(r){\boldsymbol \nabla} \times {\bf B} \right) \label{eq:induction}
\end{align}
where $\eta (r) = 1/(\mu_0 \sigma(r))$ is the \sv{magnetic} diffusivity of the medium with the conductivity taken to be a function of the radius. Since the physical parameters and flow profiles are independent of $\phi$ and $z$, we can write the magnetic field in the form of eigenmodes ${\bf B} = \widehat{\bf B} (r) \exp(i k_z z + i m \phi + \gamma t) \sv{+ c.c}$ for both the interior and the exterior region, \sv{thereby, translating the problem into an eigenvalue framework}. Substituting this form into the induction equation \eqref{eq:induction}, we get,
\begin{widetext}
\begin{align}
    \gamma \widehat{B}_r^i + \Omega(r) i m \widehat{B}_r^i + U_z(r) i k_z \widehat{B}_r^i =& - \eta(r) \left( -\frac{d}{dr} \left( \frac{1}{r} \frac{d}{dr} \left( r \widehat{B}_r^i \right) \right) + \frac{m^2}{r^2} \widehat{B}_r^i + k_z^2 \widehat{B}_r^i + \frac{2 i m}{r^2} \widehat{B}_\phi^i \right), \label{eq:Breq_in} \\
    \gamma \widehat{B}_\phi^i + \Omega(r) i m \widehat{B}_\phi^i + U\sv{_z}(r) i k_z \widehat{B}_\phi^i = & \widehat{B}_r^i r \Omega'(r) - \eta(r) \left( - \frac{d}{dr} \left( \frac{1}{r} \frac{d}{dr} \left( r \widehat{B}_\phi^i \right) \right) + \frac{m^2}{r^2} \widehat{B}_\phi^i + k_z^2 \widehat{B}_\phi^i - \frac{2 i m}{r^2} \widehat{B}_r^i \right) \nonumber \\
     & + \eta'(r) \left( \frac{1}{r} \frac{d}{dr} \left( r \widehat{B}_\phi^i \right) - \frac{im}{r} \widehat{B}_r^i  \right), \label{eq:Bpeq_in} \\
     \gamma \widehat{B}_r^o = & - \eta_0 \left( -\frac{d}{dr} \left( \frac{1}{r} \frac{d}{dr} \left( r \widehat{B}_r^{o} \right) \right) + \frac{m^2}{r^2} \widehat{B}_r^{o} + k_z^2 \widehat{B}_r^{o} + \frac{2 i m}{r^2} \widehat{B}_\phi^{o} \right), \label{eq:Breq_out} \\
     \gamma \widehat{B}_\phi^o = & - \eta_0 \left( - \frac{d}{dr} \left( \frac{1}{r} \frac{d}{dr} \left( r \widehat{B}_\phi^{o} \right) \right) + \frac{m^2}{r^2} \widehat{B}_\phi^{o} + k_z^2 \widehat{B}_\phi^{o} - \frac{2 i m}{r^2} \widehat{B}_r^{o} \right). \label{eq:Bpeq_out}
\end{align}
\end{widetext}
$\widehat{\bf B}^i$ and $\eta(r)$ denote the magnetic field and the magnetic diffusivity, respectively, in the inner domain $r \in [0, r_0]$ while for the outer domain $r \in [r_0, \infty)$, they are represented as $\widehat{\bf B}^o$ and $\eta_0$. 
\sv{For the discontinuous Ponomarenko model, we take the velocity profile to be given by constants, $U_z(r) = U_z$ and $\Omega(r) = \Omega$ that are independent of $r$ in the domain $r \leq r_0$. The rms velocity is defined by $U_{\text{rms}} = \sqrt{U_z^2 + \Omega^2 r_0^2}$. We then define the magnetic Reynolds number as $\Rm = U_{\text{rms}} r_0/\eta_0$, and the pitch of the helical motion as $\chi = U_z/(r_0 \Omega)$. Here $\eta_0=\eta(r_0)$ denotes the value of the diffusivity at $r_0$. In addition,} we consider the outer fluid to possess the conductivity of the inner fluid at $r = r_0$. Therefore, the external domain is at rest with $\eta_0= 1/(\mu_0 \sigma_0)$. For this setup, using Maxwell's equations, we obtain the magnetic boundary condition as follows,
\begin{align}
     B_r\rvert_{r_0-} = & B_r\rvert_{r_0+}; B_\phi\rvert_{r_0-} = B_\phi\rvert_{r_0+}, \label{eqn:Br_conditions} \\
     \partial_r B_r \rvert_{r_0-} = & \partial_r B_r \rvert_{r_0+}; \partial_r B_\phi \rvert_{r_0-} + \frac{r_0 \Omega}{\eta(r_0)} B_r \rvert_{r_0-} = \partial_r B_\phi \rvert_{r_0+}. \label{eqn:Bp_conditions}
\end{align}
For the magnetic field to grow exponentially in time, the real part of the growth rate $\gamma$ has to be positive. The threshold for the constant conductivity case is found to be close to $\Rm_c \approx 17.69$ obtained for critical $\chi \approx 1.31$ and $k_c \approx -0.39$, \sv{where $k_c$ denotes the vertical wavenumber corresponding to the most unstable mode}. Now, to study the effect of conductivity variation on the dynamo instability, we consider three different profiles. These conductivity profiles taken in the interior domain $r \in [0, r_0]$ are given by,
\begin{align}
    \sigma_{1,\zeta}(r) = & \frac{3 \sigma_0 }{2 + \left( \frac{r}{r_0} \right)^\zeta}, \label{eqn:profile_1} \\
    \sigma_{2,\zeta}(r) = & \frac{\left(1 + \left( \frac{r}{r_0} \right)^\zeta\right)\sigma_0}{2}, \label{eqn:profile_2}     
    \\
    \sigma_{3,\zeta}(r) = & \sigma_0 \left( 1 + \frac{1}{2} \sin \left( \zeta \pi \frac{r}{r_0} \right) \right). \label{eqn:profile_3} 
\end{align}
Here, $\zeta$ is an integer parameter that controls the strength of the conductivity variation with a larger magnitude of $\zeta$, implying a sharper variation. 
\begin{figure}
     \centering
     \begin{subfigure}[b]{0.325\textwidth}
         \centering
         \includegraphics[width=\textwidth]{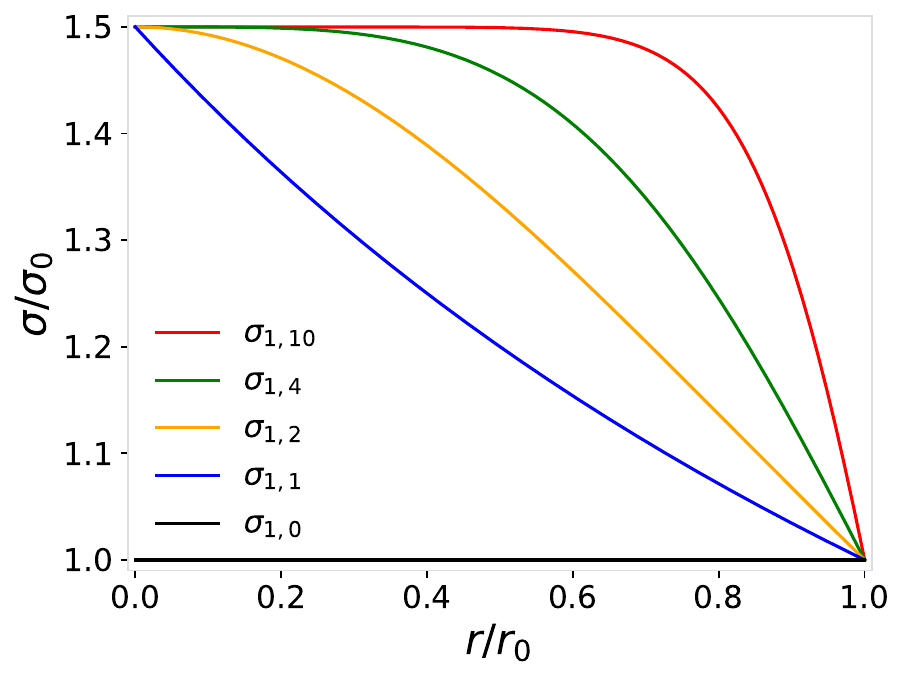}
         \caption{}
         \label{fig:flow_Re_1}
     \end{subfigure}
     \hfill
     \begin{subfigure}[b]{0.325\textwidth}
         \centering
         \includegraphics[width=\textwidth]{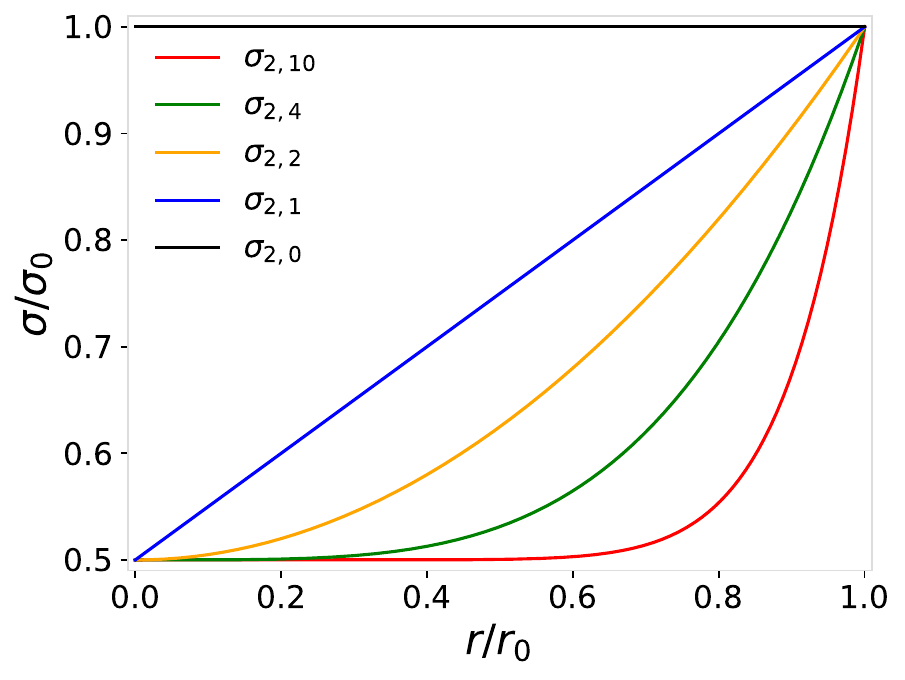}
         \caption{}
         \label{fig:flow_Re_2}
     \end{subfigure}
     \hfill
     \begin{subfigure}[b]{0.325\textwidth}
         \centering
         \includegraphics[width=\textwidth]{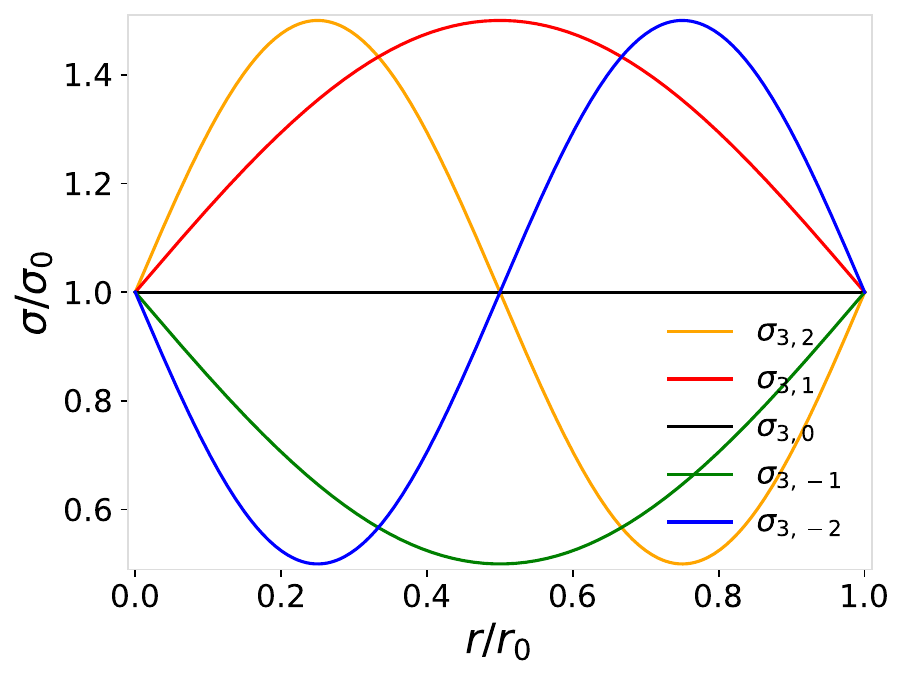}
         \caption{}
         \label{fig:flow_Re_3}
     \end{subfigure} 
        \caption{Figures show the normalised conductivity profiles in a) $\sigma_{1,\zeta}$, b) $\sigma_{2,\zeta}$ and c) $\sigma_{3, \zeta}$ as a function of the radius $r/r_0$ for different values of $\zeta$.  }
        \label{fig:flow_cond}
\end{figure}
Figure \ref{fig:flow_cond} shows the above conductivity profiles as a function of $r/r_0$ for different values of $\zeta$. The first two profiles $\sigma_{1,\zeta}$ and $\sigma_{2,\zeta}$ correspond to power law variations with monotonically decreasing/increasing conductivity from the centre $r = 0$ while $\sigma_{3, \zeta}$ varies sinusoidally. These profiles are chosen as simple models to mimic variations that have been studied earlier in spherical dynamo models \cite{wicht2019dynamo,gubbins2015core,rogers2017hottest}.

We solve the induction equation numerically for different values of $\Rm, \chi, m, k_z$ and different conductivity profiles defined in Eqs.\eqref{eqn:profile_1}, \eqref{eqn:profile_2} and \eqref{eqn:profile_3}. A central second-order finite difference scheme is used to discretise the induction equation in both domains. The inner domain is given by $r\in [0, r_0]$ while the outer domain is given by $r \in [r_0, r_{\text{out}}]$. Here $r_{\text{out}}$ is taken to be ten times larger than $r_0$ representing the boundary at infinity, we have verified that the results do not change when $r_{\text{out}}$ is increased further. The number of discretisation points $N$ is varied from \sv{$700$} at low values of $\Rm$ and $|\zeta|$ to \sv{$4000$} for large values. We take the solutions to be independent of the resolution when the results vary by less than $2\%$ as the resolution is doubled. Boundary conditions are implemented at $r = 0, r_0$ and $r_{\text{out}}$. At $r=0$, the regularity boundary conditions are implemented to obtain smooth solutions, see \cite{lewis1990physical}. We have the matching boundary conditions between the internal and the external magnetic fields, at $r= r_0$ and at $r = r_{\text{out}}$, we set the magnetic field to zero.  

\section{Threshold} \label{sec:onset} 

\subsection{Dependence on $\zeta$}
First, we study the Ponomarenko dynamo threshold for the conductivity profiles defined in Eqs. \eqref{eqn:profile_1} and \eqref{eqn:profile_2}, where the conductivity varies along the radius as a power law. The threshold is defined using the critical Reynolds number $\Rm_c$, which is given by the smallest $\Rm$ at which the growth rate of the most unstable mode becomes positive for any value of $\chi, k_z$ and $m$. 
\begin{figure}[h!]
 \includegraphics[ width=\linewidth]{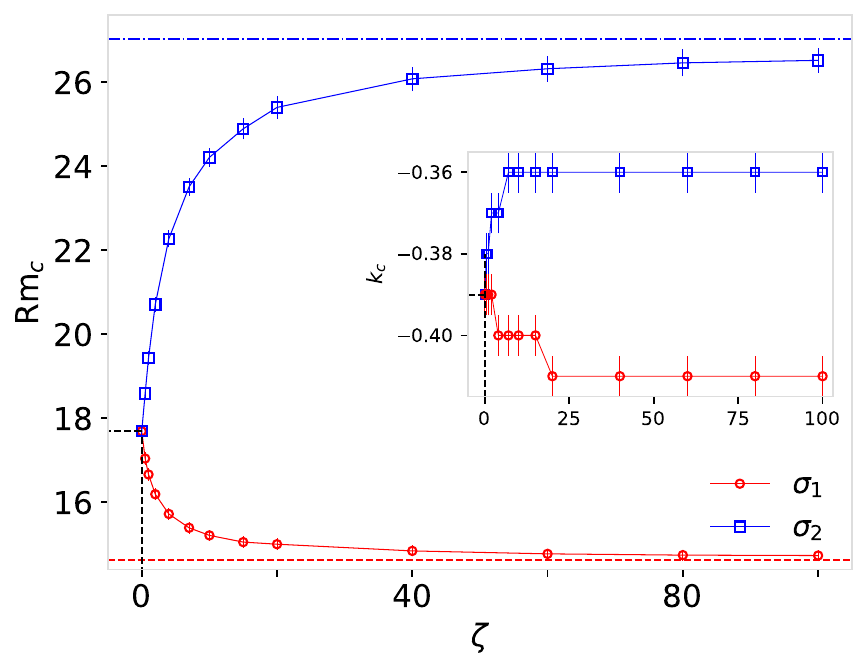}
 \caption{\label{fig:Rmc_sig12} Figure shows the critical magnetic Reynolds number $\Rm_c$ as a function of the parameter $\zeta$ for the conductivity profiles $\sigma_1$ and $\sigma_2$. The red dashed (for $\sigma_1$) and blue dash-dotted (for $\sigma_2$) lines indicate the $\Rm_c$ obtained for a jump in the conductivity across $r=r_0$ corresponding to the asymptote $\zeta \rightarrow \infty$. The inset shows the vertical wavenumber corresponding to the most
unstable mode.}
\end{figure}
Figure \ref{fig:Rmc_sig12} shows the variation of $\Rm_c$ for the power law conductivity profiles as a function of the parameter $\zeta$. For the $\sigma_1$ profiles, we find that the threshold decreases when more of the inner domain is at a higher conductivity as compared to the constant conductivity case, i.e., when the mean value of $\sigma(r)$ in the inner domain becomes larger. Similarly, for $\sigma_2$ profiles, the threshold increases as more of the interior domain has lower conductivity when compared to the constant conductivity case. The asymptote of the curves can be found by considering the discrete Ponomarenko problem with constant conductivity in all the regions and a jump in the conductivity value across $r =r_0$. The details of this derivation and the jump conductivity problem are discussed in detail in Appendix \ref{App:jump_cond}. In the large $\zeta$ limit, $\Rm_c$ asymptotes to a value independent of $\zeta$ given by the red dashed (for $\sigma_1$) and blue dash-dotted (for $\sigma_2$) lines in Fig. \ref{fig:Rmc_sig12}. The value of the asymptote is controlled by the conductivity ratio between the inner and the outer domains. 

The vertical wavenumber corresponding to the most unstable mode, $k_c$, is shown in the inset of Fig. \ref{fig:Rmc_sig12}. The wavenumber of the unstable mode decreases for the profile $\sigma_{1, \zeta}$ and increases for the profile $\sigma_{2,\zeta}$ as $\zeta$ is increased before saturating to a $\zeta$ independent value. \sv{An opposite} behaviour is found for the critical $\chi$ at which the dynamo threshold is found. \sv{It increases for the profile $\sigma_{1, \zeta}$ and decreases for the profile $\sigma_{2,\zeta}$ as $\zeta$ is increased}, going to a $\zeta$ independent asymptote at large $\zeta$. \sv{This follows from the resonance condition $m\Omega + k_zU_z = 0$, where we expect that $\chi$ varies as the negative inverse of $k_z$}.
\begin{figure}[h!]
 \includegraphics[width=\linewidth]{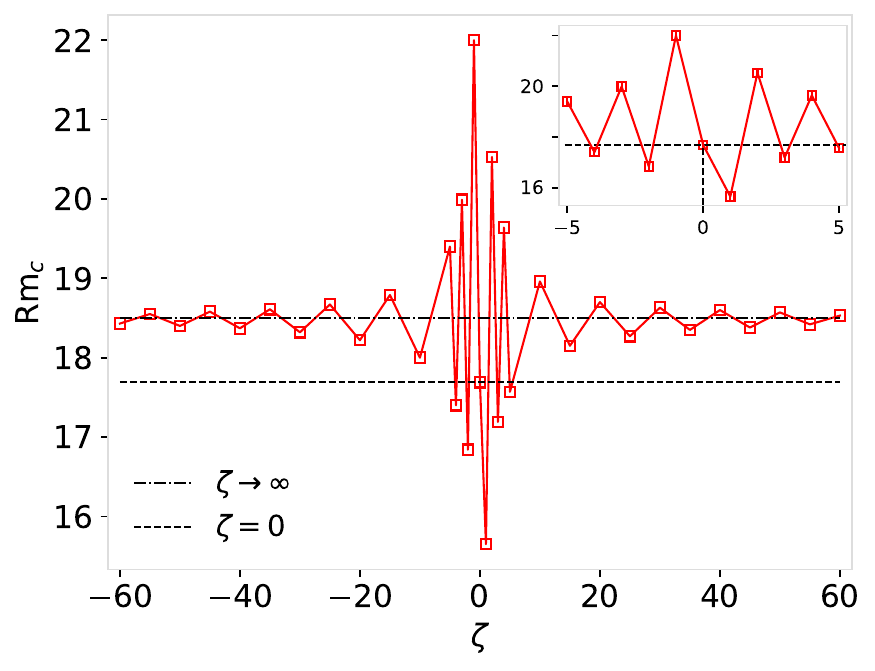}
 \caption{\label{fig:Rmc_sig3} Figure shows the critical magnetic Reynolds number $\Rm_c$ as a function of the parameter $\zeta$ for the sinusoidal conductivity profile $\sigma_3$. The dashed line indicates the threshold of the constant conductivity case. \sv{The inset shows the fluctuation of $\Rm_c$ near the threshold of the constant conductivity case for small $\zeta$. } }
\end{figure}

For the sinusoidal conductivity profiles $\sigma_{3,\zeta}$, Figure \ref{fig:Rmc_sig3} shows the critical magnetic Reynolds number as a function of $\zeta$. In the cases of $\zeta = 1$ and $ -1$, $\Rm_c$ changes following the decrease/increase of the mean conductivity in the inner domain. For the remaining values of $\zeta$, we find that $\Rm_c$ fluctuates near the threshold of the constant conductivity case such that the value increases when the conductivity is decreased closer to $r = r_0$ while in the case of increased conductivity near $r_0$, the threshold decreases. This explains the increase in $\Rm_c$ for $\zeta = 2, -3, 4$ and the decrease in $\Rm_c$ for $\zeta = -2, 3, -4$, \sv{as seen in the inset of Fig. \ref{fig:Rmc_sig3}}. For even larger values of $\zeta$, $\Rm_c$ asymptotes to a value larger than the constant conductivity case. This can be explained by the averaging techniques using which we find the effective diffusivity due to the rapidly oscillating conductivity, leading to an increase in the threshold, see Appendix A. We see from the figures \ref{fig:Rmc_sig12} and \ref{fig:Rmc_sig3} that the $\Rm_c$ varies strongly with $\zeta$ for $|\zeta| \lesssim 10$ after which it tends to a $\zeta$ independent asymptote.

\subsection{Ohmic Dissipation}

\begin{figure}[h!]
 \includegraphics[width=\linewidth]{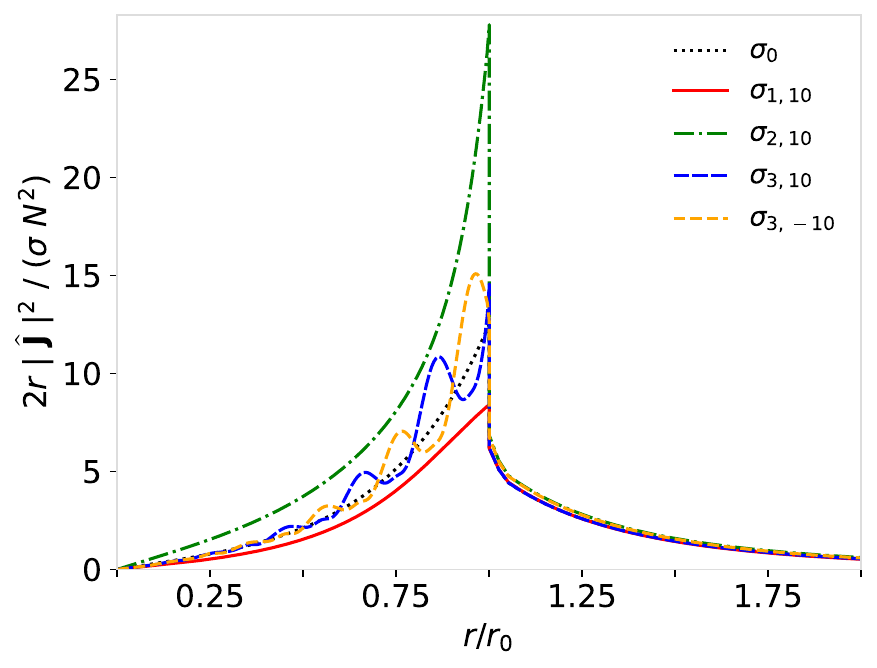}
 \caption{\label{fig:eigenmodes} Figure shows the normalised dissipation density $2r|\widehat{\bf J}|^2/(\sigma N^2)$ as a function of $r/r_0$ for different conductivity profiles as labeled.  }
\end{figure}

To quantify the change in $\Rm_c$, we look at the dissipation rate and its dependence on the form of the conductivity profile. \sv{A measure of the ohmic dissipation, normalised with magnetic energy, is given as,
\begin{equation}\label{ohmDissbar}
    \bar{D} = \frac{\int |{\boldsymbol \nabla} \times {\bf B}|^2/(\mu_0\sigma )\mbox{ }dV}{\int |{\bf B}|^2 dV}
\end{equation}}
\sv{where the integral is over the volume V of the fluid inside the cylinder and ${\mu_0}$ is the magnetic permeability. 
Expressing the magnetic field in terms of eigenmodes as ${\bf B} = \widehat{\bf B} (r) \exp(i k_z z + i m \phi + \gamma t) + c.c$, and following \cite{kaiser1999kinematic}, we get,
\begin{equation}
    \bar{D} = \frac{1}{\int\frac{ 2|{\bf \widehat{B}}|^2}{\mu_0} dV} \iiint \left(\frac{2|{\bf \widehat{J}}|^2 r}{\sigma }\right) dr d\phi dz
\end{equation}
where current density $\widehat{\bf J} (r)$ for the mode $\widehat{\bf B}$ is taken to be $\widehat{\bf J} (r) = (1/\mu_0)\left( {\bf e}_r \partial_r + {\bf e}_\phi i m/r + {\bf e}_z i k_z \right) \times \widehat{\bf B}$. Therefore, the measure of ohmic dissipation can be expressed as,
\begin{equation}
     \bar{D}=\frac{1}{V}\iiint \left(\frac{2r|{\bf \widehat{J}}|^2}{\sigma N^2}\right) dr d\phi dz
    \label{eqn:dbar}  
\end{equation}
where $N^2$ is the normalization factor given by magnetic energy density in the inner domain, defined as $N^2 = (1/V) \iiint 2|{\bf \widehat{B}}|^2r/\mu_0  dr d\phi dz$. Figure \ref{fig:eigenmodes} shows the normalised dissipation density as a function of $r/r_0$ for different conductivity profiles.} We find that there is an increase/decrease of the threshold of the dynamo $\Rm_c$ when the dissipation rate is larger/smaller than the constant conductivity case $\sigma_0$ (dotted line). For the cases with $\sigma_{3,10}$ and $\sigma_{3,-10}$, the eigenmodes are found to oscillate with $r$, which is also seen in the dissipation rates shown in Fig. \ref{fig:eigenmodes}.


\section{Large $\Rm$ limit} \label{sec:largeRm}

\begin{figure}[h!]
     \begin{subfigure}[h]{0.475\textwidth}
 \includegraphics[width=0.95\linewidth]{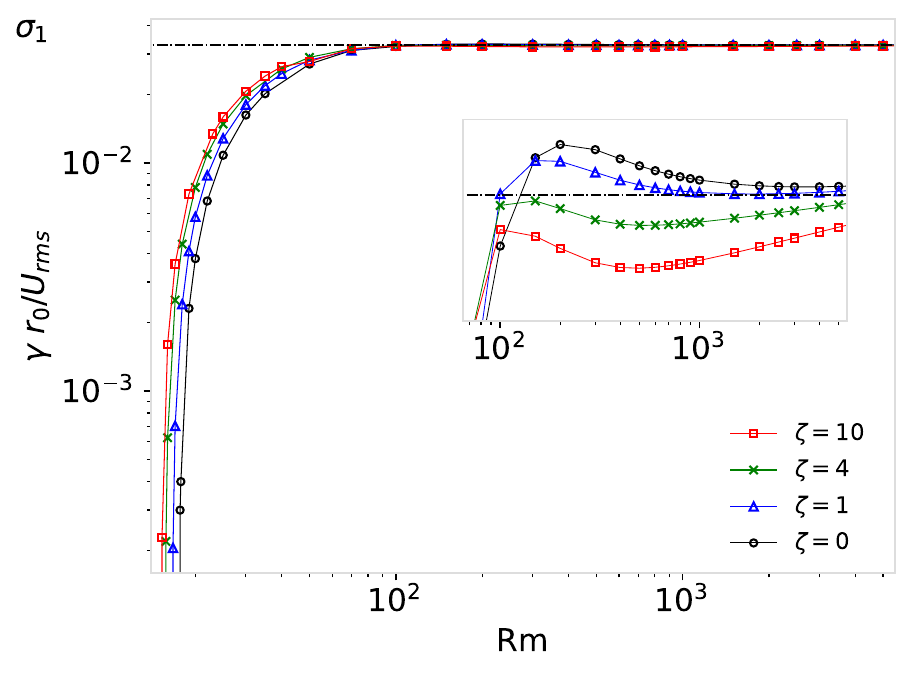}
  \caption{\label{fig:gam_Rm_sig1}}
      \end{subfigure}
     \begin{subfigure}[h]{0.475\textwidth}
 \includegraphics[width=0.95\linewidth]{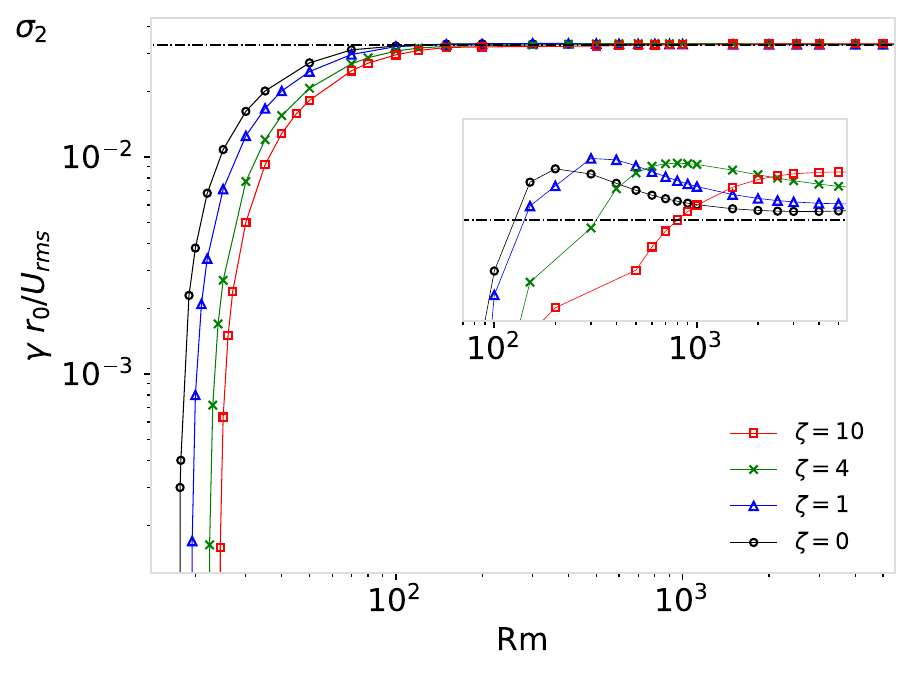}
  \caption{\label{fig:gam_Rm_sig2}}
      \end{subfigure}
      \begin{subfigure}[h]{0.475\textwidth}
 \includegraphics[width=0.95\linewidth]{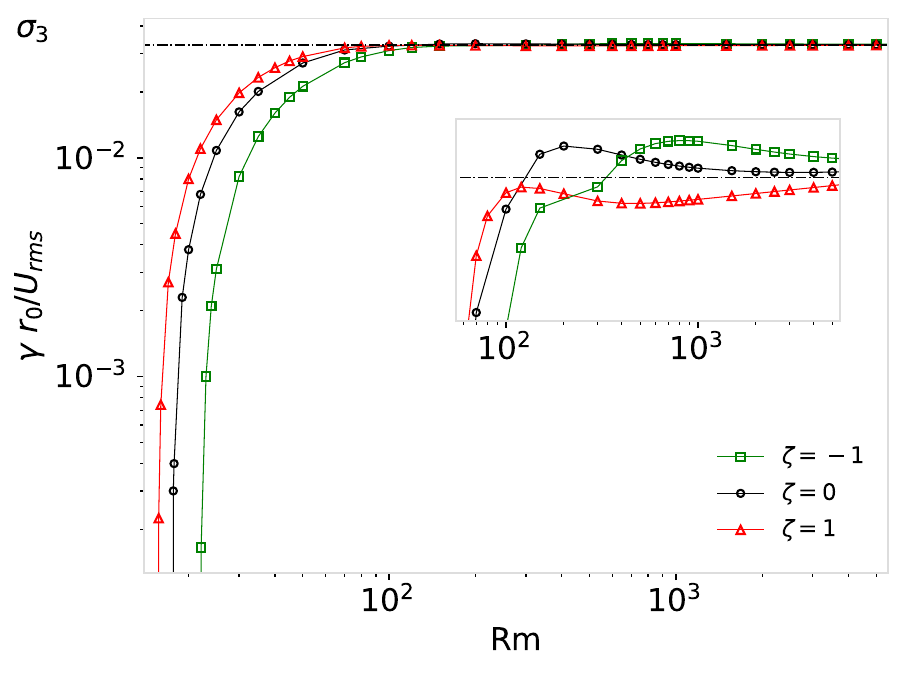}
  \caption{\label{fig:gam_Rm_sig3}}
      \end{subfigure}
 \caption{\label{fig:large_Rm_limit} Figures show the large $\Rm$ behaviour of the \sv{real value of the} growth rate $\gamma$ for conductivity profiles, a) $\sigma_1$, b) $\sigma_2$ and c) $\sigma_3$ for different values of $\zeta$. The inset shows \sv{a crossover of the curves at moderate $\Rm$, and the dash-dot line represents the theoretical real value of the growth rate for the constant conductivity case}.}
\end{figure}

Next, we look at the system far away from the threshold. \sv{Here, for the most unstable mode we have $m$ and $k_z$ $\propto \Rm^{1/2}$}. The \sv{real value of the} maximum growth rate $\gamma$ is shown as a function of $\Rm$ for a given $\chi = 1.31$ in Figs. \ref{fig:gam_Rm_sig1}, \ref{fig:gam_Rm_sig2} and \ref{fig:gam_Rm_sig3} for the conductivity profiles $\sigma_1, \sigma_2$ and $\sigma_3$, respectively. \sv{The growth rate is expressed in units of advective time scales, hence, a factor of $r_0/U_{rms}$ is multiplied}. While the critical Reynolds number $\Rm_c$ is found to depend on the value of $\zeta$ as described by Figs. \ref{fig:Rmc_sig12} and \ref{fig:Rmc_sig3}, but for larger values of $\Rm$, we find that the growth rate $\gamma$ for different values of $\zeta$ start merging into the solution given by the constant conductivity case where $\zeta = 0$. \sv{This can be seen in Figs. \ref{fig:gam_Rm_sig1}, \ref{fig:gam_Rm_sig2} and \ref{fig:gam_Rm_sig3} where the dash-dot line is representative of the theoretical real value of the growth rate for $\zeta = 0$.} Thus, in the large $\Rm$ limit, the conductivity variations are sub-dominant for the finite $\zeta$ cases. We can find the leading order correction to the growth rate $\gamma$ at large $\Rm$ analytically by making use of the fact that the magnetic field becomes concentrated in a thin region near $r = r_0$ of thickness proportional to $\Rm^{-1/2}$. Following \cite{gilbert1988fast}, we perform an asymptotic analysis whose details are mentioned in Appendix \ref{App:asymptotics}. We find that the correction to the growth rate due to conductivity variation is given by,
    \begin{align}
        \delta \gamma = -\left. \frac{\eta'}{\eta} \right|_{r = r_0} \frac{4 \gamma_0 - 3 \alpha^2 \eta\rvert_{r = r_0}}{12 \alpha} \label{eqn:gamma_corr}
    \end{align}
    where $\delta \gamma$ is the leading order correction\sv{, modifying both the real and imaginary parts of the growth rate}, with $\alpha$ and $\gamma_0$ defined in Appendix \ref{App:asymptotics}. The first term on the right-hand side of the Eq. \eqref{eqn:gamma_corr} is proportional to the slope of the diffusivity profile at $r = r_0$, implying that a steeper variation of the conductivity at $r = r_0$ leads to a larger deviation from the case of no conductivity variation. For the conductivity profiles used, given in Eqs. \eqref{eqn:profile_1}, \eqref{eqn:profile_2} and \eqref{eqn:profile_3}, the slope of the diffusivity profile at $r = r_0$, $\eta'/\eta \rvert_{r =r_0} \propto \zeta$, indicating that the correction is given by $\delta \gamma \propto \zeta$, which is also seen in the numerical results shown in Figs. \ref{fig:gam_Rm_sig1}, \ref{fig:gam_Rm_sig2} and \ref{fig:gam_Rm_sig3} where larger values of $\zeta$ deviate further from the asymptote for a given $\Rm$.  The second term is proportional to $\Rm^{-1/2}$, which indicates that the correction to the growth rate is at the sub-leading order $\Rm^{-1/2}$, also seen in the numerical solution, refer to Fig. \ref{fig:deviation} in Appendix \ref{App:asymptotics}. 
    

Furthermore, the sign of the modification in the growth rate $\delta \gamma$ is determined only by the term $\eta'(r)$ as the term $(4\gamma_0 - 3 \alpha^2 \eta)/(12 \alpha)$ is found to be positive definite, see Appendix \ref{App:asymptotics}. We find a positive correction to the growth rate when the diffusivity profile $\eta(r)$ has a negative slope at $r = r_0$, and $\delta \gamma$ is negative when $\eta(r)$ has a positive slope at $r = r_0$. Given that $\sigma(r)$ is the inverse of $\eta(r)$, from the profiles shown in Eqs. \eqref{eqn:profile_1}, \eqref{eqn:profile_2} and \eqref{eqn:profile_3}, we find that for $\sigma_{1,\zeta}$ case, larger $\zeta$ leads to a smaller effective growth while for $\sigma_{2,\zeta}$, larger $\zeta$ leads to a larger effective growth. In the case of the sinusoidal profile $\sigma_{3,\zeta}$, using the expression \eqref{eqn:gamma_corr} we find that the modification $\delta \gamma$ is positive when $\zeta \cos (\zeta \pi)$ is positive while $\delta \gamma$ is negative when $\zeta \cos(\zeta \pi)$ is negative. This is also seen in the inset of the figures \ref{fig:gam_Rm_sig1}, \ref{fig:gam_Rm_sig2} and \ref{fig:gam_Rm_sig3} with the numerical solutions following the asymptotic prediction of Eq.\eqref{eqn:gamma_corr}.

The inset also shows a cross-over regime at intermediate values of $\Rm$ where the growth rates curves for different $\zeta$ reverse their ordering. For a particular $\zeta$ curve, if at lower values of $\Rm$, the growth rate is found to be larger/smaller than the $\zeta = 0$ case, then, at larger values of $\Rm$, it becomes smaller/larger than the $\zeta = 0$ case. This is seen for all the profiles $\sigma_{1, \zeta}, \sigma_{2, \zeta}$ and $ \sigma_{3, \zeta}$ shown in Figs. \ref{fig:gam_Rm_sig1}, \ref{fig:gam_Rm_sig2} and \ref{fig:gam_Rm_sig3}. This behaviour can be understood from the conductivity profiles and from the structure of the unstable mode. For the low Rm regime, the magnetic field decays slowly as we move away from $r_0$. Therefore, a larger/smaller $\sigma$ compared to $\sigma_0$ near the boundary $r_0$ leads to a larger/smaller growth rate, implying higher growth rates for profiles that have $\eta'(r)$ positive at $r = r_0$, seen in Figs. \ref{fig:gam_Rm_sig1}, \ref{fig:gam_Rm_sig2} and \ref{fig:gam_Rm_sig3}. While for larger values of $\Rm$, the magnetic field is concentrated in a thin region near $r_0$ of width proportional to $\Rm^{-1/2}$, with the behaviour of $\sigma$ at $r = r_0$ found to determine the sign of the growth rate correction which predicts a larger growth rate for negative values of $\eta'(r)$ at $r = r_0$. \sv{A similar behaviour is observed for the imaginary component of the growth rate, and consequently, the inferences remain the same.} Thus, for the conductivity profiles studied here, depending upon the sign of $\eta'(r)$, the ordering of the growth rates changes at low and large $\Rm$, leading to a cross-over regime at intermediate values. Hence, the same conductivity profile can give an increase or a decrease in growth rate depending on Rm \sv{as compared to the constant conductivity case}.

We can, therefore, say that in the large Rm limit, a negative sign of $\eta'(r)$ at $r_0$  aids the dynamo instability. To be precise, at large enough Rm values, when both $\eta'(r)$ and $\Omega'(r)$ have the same sign near $r = r_0$, the variable conductivity effect along with the $\Omega$ effect converts $B_r$ to $B_{\phi}$ while the conversion from $B_\phi$ to $B_r$ occurs through diffusion. Similar effects have been observed in other scenarios where conductivity variations correlated with vorticity fluctuations aid in the dynamo effect \cite{petrelis2016fluctuations}.

\section{Effect of magnetic boundary conditions}\label{sec:mbc}

In this section, we investigate the effect of radial conductivity variations in the presence of different magnetic boundary conditions and with a realistic velocity profile. \sv{Reinstating the radial variation in the velocity field}, we consider a velocity profile ${\bf u} (r)$ of the conducting fluid in the domain $[0, r_0]$, given by, 
\begin{align}
    {\bf u} =  r \Omega_0 \left( 1 - \frac{r^2}{r_0^2} \right) {\bf e}_\phi + U_0 \left( 1 - \frac{r^2}{r_0^2} \right) {\bf e}_z.
\end{align}
Similar velocity profiles have been used in other studies \cite{peyrot2007parametric,peyrot2008oscillating}. The growth rate of the magnetic field tends to zero in the infinite $\Rm$ limit \cite{gilbert1988fast} due to the finite shear as compared with the infinite shear present in the discontinuous velocity profile. We define the rms velocity scale as $U = \sqrt{ U_0^2 + r_0^2 \Omega_0^2}$. For $r > r_0$, we consider different magnetic boundary conditions to see the effect of variable conductivity profiles. We consider the following boundary conditions, a) Same fluid at rest, b) Insulator, c) Perfect conductor and d) Pseudo vacuum/Ferromagnetic boundary conditions. The same fluid and insulator require the external fields to be solved and matched with the internal fields, while for the cases of perfect conductor and pseudo vacuum, the boundary conditions are locally imposed at $r = r_0$. The mathematical constraints imposed for these boundary conditions can be found in \cite{moffatt2019self}. 

\begin{figure}[h!]
     \begin{subfigure}[h]{0.47\textwidth}
 \includegraphics[width=\linewidth]{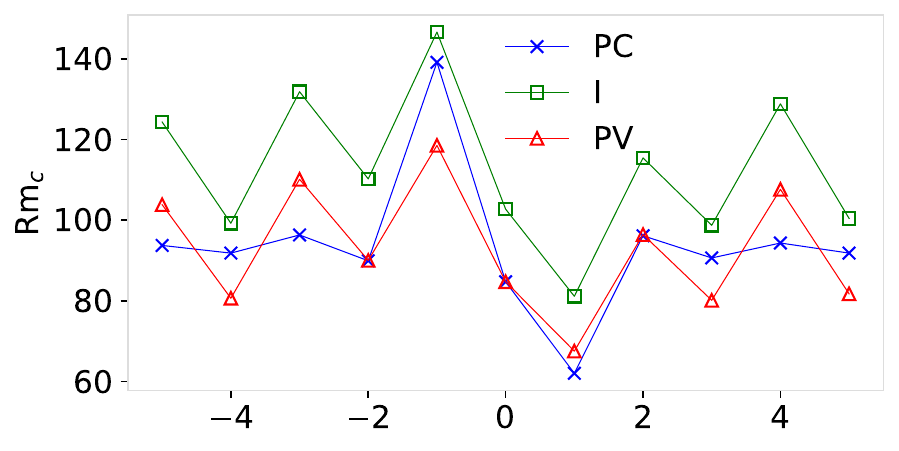}
  \caption{\label{fig:bound_var_1}}
      \end{subfigure}
     \begin{subfigure}[h]{0.47\textwidth}
 \includegraphics[width=\linewidth]{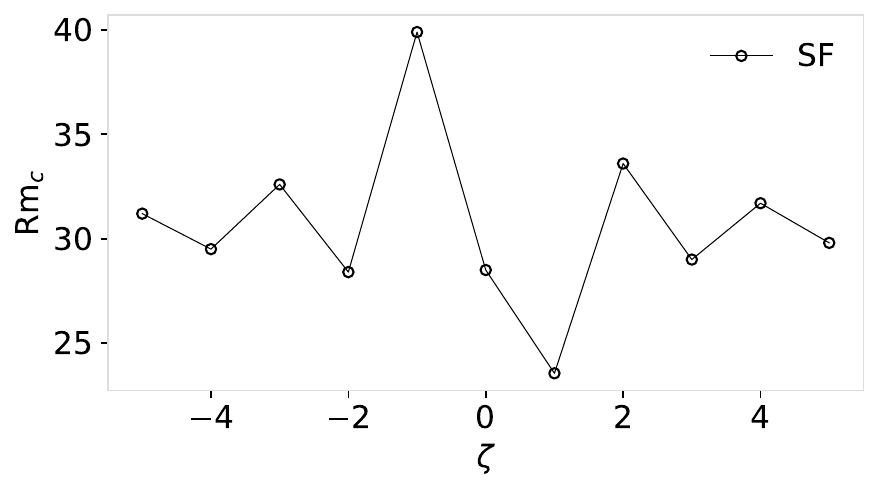}
  \caption{\label{fig:bound_var_2}}
      \end{subfigure}
 \caption{\label{fig:diff_bcs} Figures show the critical Reynolds numbers $\Rm_c$ as a function of $\zeta$ for the sinusoidal conductivity profiles. The different curves correspond to the magnetic boundary conditions in a) for the perfect conductor (PC), insulator (I), pseudo vacuum (PV), and b) for the same fluid (SF). }
\end{figure}

 Figs. \ref{fig:bound_var_1} and \ref{fig:bound_var_2} show the critical magnetic Reynolds number $\Rm_c$ as a function of $\zeta$ parameter for the sinusoidal conductivity profiles. Fig. \ref{fig:bound_var_1} shows the critical Reynolds number $\Rm_c$ for the perfect conductor (PC), insulator (I) and pseudo vacuum (PV). Their $\Rm_c$ is much larger than the $\Rm_c$ for the same fluid case shown in Fig. \ref{fig:bound_var_2}. As seen from these figures, the conductivity variation shows similar changes in the $\Rm_c$ for the different magnetic boundary conditions that have been explored. The insulator boundary condition shows the largest $\Rm_c$ while perfect conductor and pseudo vacuum show similar values of $\Rm_c$. At $\zeta = 0$, both PC and PV have the same threshold due to the equivalence between the two problems as shown in \sv{\cite{marcotte2021enhanced,favier2013growth}}. The threshold behaviour for the other conductivity profiles $\sigma_{1,\zeta} $ and $\sigma_{2,\zeta}$ follow similar trends and are not shown here. We find that the modifications due to radial conductivity variations are independent of the velocity profile and the magnetic boundary conditions that were used.

\section{Conclusion} \label{sec:Conclusion}
In this study, we have quantified the dynamo instability in the Ponomarenko dynamo with radially varying conductivity. For monotonically increasing/decreasing profiles, we find an increase/decrease in the threshold, which asymptotes to a constant in the limit of sharp variation. For rapidly varying sinusoidal profiles, $\Rm_c$ is found to be larger than the constant conductivity case due to modified dissipation arising from the rapid variations. Similar behaviour is also found in the case of a smooth velocity profile and for different magnetic boundary conditions. It is interesting to note that the same fluid (SF) boundary condition has the lowest $\Rm_c$. Further, in the large $\Rm$ limit, the leading order correction to the growth rate is found to be determined by $\eta'/\eta \rvert_{r =r_0} $. At large Rm, an increase/decrease in growth rate is found when the conductivity \sv{increases/decreases} close to the discontinuity, while an exact opposite behaviour is seen near the threshold, indicating a crossover at intermediate Rm values. 

With a very simple flow, we are able to explicitly illustrate the effects of the gradient of conductivity on the dynamo instability. It is to be noted that these effects are independent of the flow field. The relative increase/decrease of the $\Rm_c$ is a measure of the effective diffusivity in the variable conductivity model, also seen in atmospheric dynamo models where ohmic dissipation is modified due to variations in conductivity \cite{rogers2017hottest}. Although the correction term in the growth rate due to radial variations is at the sub-leading order, in large Rm regime where most astrophysical systems operate, we find that a negative $\eta'$ and $\Omega'$ or a positive $\sigma'$ and a negative $\Omega'$ aids the dynamo instability in regenerating B$_{\phi}$ from B$_r$, similar to other observations \cite{rogers2017hottest,petrelis2016fluctuations}. Results from this theoretical model indicate that conductivity variations affect the dynamo growth rate when $\eta' \Rm^{-1/2}$ is large. \sv{On the whole, the validity of our results stands in regimes where convection/mixing is not strong enough to alter the background profiles}, and it could be of importance while modelling astrophysical systems.

We have looked into the kinematic phase, and a similar study can be done for the saturation phase \sv{\cite{nunez2001saturation, dobler2002nonlinear}}. Attempts have already been made to explain the zonal flow depth in gas giants using the variable conductivity dynamo models wherein the Coriolis and the Lorentz forces balance the magnetic diffusion \cite{heimpel2011relationship}. More systematic studies can be done in the spherical domain to see in what way the results in our work can be translated to better explain the balance required between the flow strength and the field amplitude for the generation of the dipole-dominated dynamos that are observed in gas giants \cite{duarte2018physical}. Further, adding radial variation in conductivity while modelling hot Jupiters may explain the high value of inflated radii in these systems, which has not been achieved so far \cite{rogers2017hottest}. Moreover, we have considered a laminar flow, whereas turbulence may enhance diffusion even further, and it could be of interest to see how this changes the effects of conductivity variation in the dynamo process.

 
\begin{acknowledgments}
SV acknowledges the support from the Prime Minister’s Research Fellows (PMRF) scheme, Ministry of Education, Govt. of India. The authors would like to thank the computing resources and support provided by PARAM Shakti, the supercomputing facility of IIT Kharagpur, established under the National Supercomputing Mission (NSM), Government of India and supported by Centre for Development of Advanced Computing (CDAC), Pune. We also acknowledge support from NSM Grant No. DST/NSM/R\&D HPC Applications/2021/03.11, and the Start-up Research Grant No. SRG/2021/001229 from Science \& Engineering Research Board (SERB), India.
\end{acknowledgments}
\appendix

\section{Large $\zeta$ limit} \label{App:jump_cond}

\begin{figure}[h!]
 \centering
 \includegraphics[width=\linewidth]{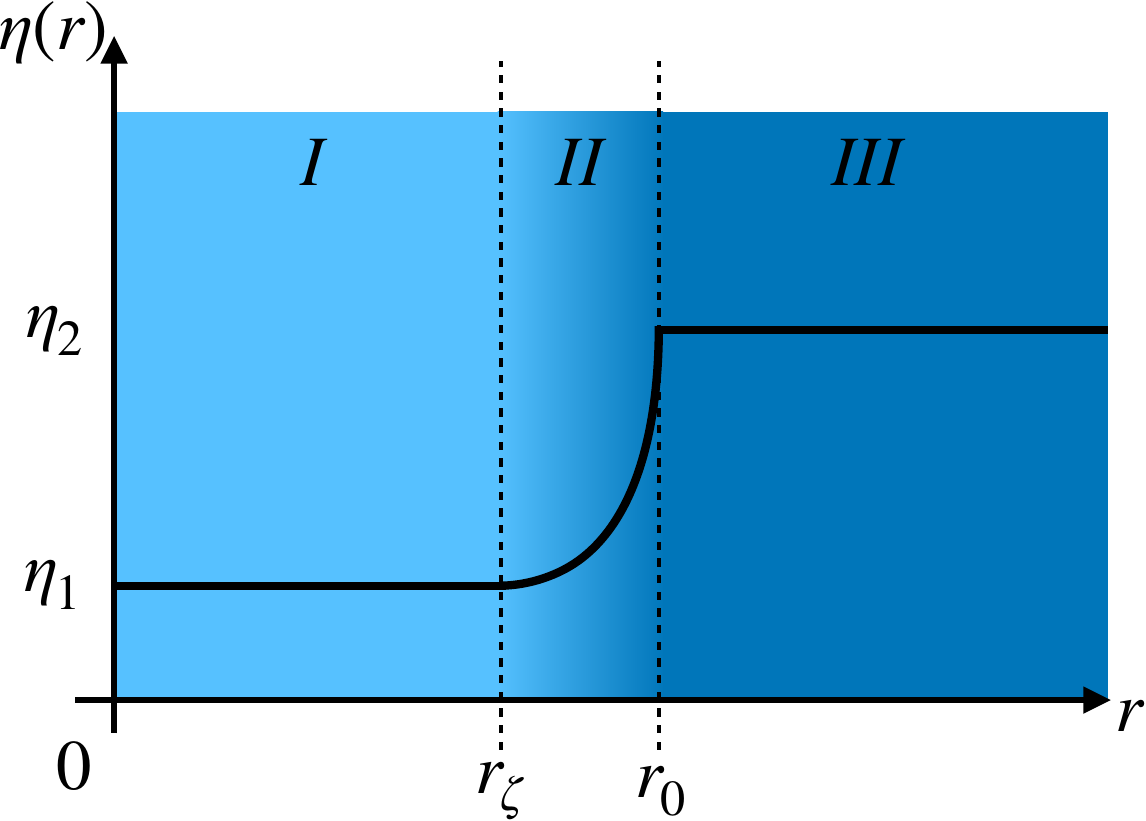}
 \caption{\label{fig:profile_eta} Figure shows an example of a monotonically increasing diffusivity function $\eta$ with the variation being concentrated near $r_0$. }
\end{figure}

In this section, we find the dynamo threshold in the large $\zeta$ limit of the conductivity profiles  $\sigma_{1,\zeta} (r)$ and $ \sigma_{2,\zeta} (r)$. We consider a general monotonically increasing or decreasing diffusivity profile, which varies sharply near $r = r_0$. Fig. \ref{fig:profile_eta} shows an example of an increasing $\eta$ function, where the diffusivity varies sharply in a region $[r_\zeta, r_0]$ with $r_\zeta$ as the radius at which the diffusivity variation begins to be significant. If we take the diffusivity at $r= r_\zeta$ to be one percent \sv{larger/smaller} than the diffusivity value at $0$, it gives \sv{ $r_\zeta = r_0/50^{1/\zeta}$} for the profile $\sigma_{1,\zeta}$ and \sv{$r_\zeta = r_0/99^{1/\zeta}$ }for $\sigma_{2,\zeta}$. Thus, in the limit $\zeta \rightarrow \infty$, $r_\zeta \rightarrow r_0$. We now split the whole domain into three regions such that $[0, r_\zeta]$ is denoted as region I, $[r_\zeta, r_0]$ with strong diffusivity variation is denoted as region II and $[r_0, \infty)$ where the fluid is stationary is the region III. Note that the diffusivity (velocity) is considered constant in regions I and III (I and II),  while the variation in diffusivity is concentrated only in region II.

\noindent
Next, we solve for the magnetic field in the three regions separately and match it across the boundaries. We denote the field (its components) as ${\bf B}^{(1)} (B_r^{(1)}, B_\phi^{(1)})$, ${\bf B}^{(2)}(B_r^{(2)}, B_\phi^{(2)})$ and ${\bf B}^{(3)}(B_r^{(3)}, B_\phi^{(3)})$ corresponding to the regions I, II and III, respectively.  Since the \sv{diffusivity} and the velocity field are continuous across I and II, the magnetic boundary conditions can be written by the continuity of fields and its first derivatives as,
\begin{align}
     B_r^{(1)}\rvert_{r_0} = & B_r^{(2)}\rvert_{r_0}; B_\phi^{(1)}\rvert_{r_0} = B_\phi^{(2)}\rvert_{r_0}, \label{eqn:12_Br_conditions} \\
     \partial_r B_r^{(1)} \rvert_{r_0} = & \partial_r B_r^{(2)} \rvert_{r_0}; \partial_r  B_\phi^{(1)} \rvert_{r_0} =  \partial_r  B_\phi^{(2)} \rvert{r_0}. \label{eqn:12_Bp_conditions}
\end{align}
Across II and III, the magnetic boundary conditions are the same as the standard DPD problem where the velocity field undergoes a discontinuity,
\begin{align}
     B_r^{(2)}\rvert_{r_0} = & B_r^{(3)}\rvert_{r_0}; B_\phi^{(2)}\rvert_{r_0} = B_\phi^{(3)}\rvert_{r_0}, \label{eqn:23_Br_conditions} \\
     \partial_r B_r^{(2)} \rvert_{r_0} = & \partial_r B_r^{(3)} \rvert_{r_0}; \partial_r B_\phi^{(2)} \rvert_{r_0} + \frac{r_0 \Omega}{\eta(r_0)} B_r^{(2)} \rvert_{r_0} \nonumber \\ 
       = \partial_r B_\phi^{(3)} \rvert_{r_0}. \label{eqn:23_Bp_conditions}
\end{align}
Finally, in region II, we solve for the field ${\bf B}^{(2)}$ as $r_\zeta \rightarrow r_0$ in the large $\zeta$ limit. The equation for $B_r$, Eq.(\ref{eq:Breq_in}), does not contain any term proportional to $\eta'(r)$ \sv{implying that} as $r_\zeta \rightarrow r_0$ we have both $B_r$ and $\partial_r B_r$ to be continuous. In the governing equation for $B_\phi$, Eq.(\ref{eq:Bpeq_in}), the term corresponding to $\eta'(r)$ will be the most dominant term with $\eta'(r) \rightarrow \infty$ as $r_\zeta \rightarrow r_0$. Therefore, to solve the equation, a jump condition is needed in the radial derivative of $B_\phi$ to balance the terms proportional to $\eta'(r)$. Integrating the Eq.(\ref{eq:Bpeq_in}) in $r$ and keeping terms that are proportional to $\eta'$ and the derivatives of $B_\phi$, we end up with the jump condition across the region II as,
\begin{align}
\left[ \eta \partial_r B_\phi^{(2)} \right]_{r_\zeta}^{r_0} = \left( \eta_2 - \eta_1 \right) \left. \left(- \frac{B_\phi^{(2)}}{r} + \frac{1}{r} \partial_\phi B_r^{(2)} \right)\right|_{r_0} , \label{eq:jump_cond}
\end{align} 
where the symbol $\left[ \; \right]_{r_\zeta}^{r_0}$ denotes the jump condition across region II, and the quantity on the right-hand side is evaluated inside the cylinder at $r_0$. Except for $\partial_r B_\phi$ which follows the jump condition shown in Eq. \eqref{eq:jump_cond}, the other quantities such as $B_r, B_\phi, \partial_r B_r$ are taken to be \sv{continuous} across region II as $r_\zeta \rightarrow r_0$. Hence, in the limit $\zeta \rightarrow \infty$ and using Eq. \eqref{eq:jump_cond}, we can write the solutions to the magnetic fields in region II in terms of the region I as, 
\begin{align}
B_r^{(2)}\rvert_{r_0} = & B_r^{(1)}\rvert_{r_\zeta}; B_\phi^{(2)}\rvert_{r_0} = B_\phi^{(1)}\rvert_{r_\zeta}, \label{eqn:regionII_Br_sol} \\
\partial_r B_r^{(2)} \rvert_{r_0} = & \partial_r B_r^{(1)} \rvert_{r_\zeta}; \eta_2 \left. \partial_r B_\phi^{(2)} \right|_{r_0} - \eta_1 \left. \partial_r B_\phi^{(1)} \right|_{r_\zeta} = \nonumber \\
\left( \eta_2 - \eta_1 \right) &\left. \left(- \frac{B_\phi^{(2)}}{r} + \frac{1}{r} \partial_\phi B_r^{(2)} \right)\right|_{r_0} , \label{eq:regionII_sol}
\end{align} 
Taking the $\zeta \rightarrow \infty$ limit and substituting the relations in Eqs.\eqref{eqn:regionII_Br_sol} and \eqref{eq:regionII_sol} into Eqs.\eqref{eqn:23_Br_conditions} and \eqref{eqn:23_Bp_conditions}, we get the boundary conditions relating the magnetic field in region I and region III as,
\begin{align}
     B_r^{(1)}\rvert_{r_0} & = B_r^{(3)}\rvert_{r_0}; B_\phi^{(1)}\rvert_{r_0} = B_\phi^{(3)}\rvert_{r_0},  \label{eqn:jump_Br_conditions} \\
         \nonumber \\ 
       \partial_r B_r^{(1)} \rvert_{r_0} & = \partial_r B_r^{(3)} \rvert_{r_0}; \frac{\eta_1}{r} \left. \left( \partial_r \left( r B_\phi^{(1)} \right)  - \partial_\phi B_r^{(1)} \right)\right|_{r_0} \nonumber \\
        + r \Omega B_r^{(1)} \rvert_{r_0}  & = \frac{\eta_2}{r} \left. \left( \partial_r  \left( r B_\phi^{(3)} \right) - \partial_{\phi} B_r^{(3)} \right)\right|_{r_0}. \label{eqn:jump_Bp_conditions}
\end{align}
The \eqref{eqn:jump_Br_conditions} and \eqref{eqn:jump_Bp_conditions} boundary conditions correspond to the problem where both the conductivity and the velocity profile are discontinuous across $r_0$. Thus, we conclude that in the limit of sharp conductivity variation, $\zeta \rightarrow \infty$, the dynamo problem can be studied using a conductivity jump profile.

Next, we look for the large $\zeta$ asymptote for the sinusoidal profile $\sigma_{3, \zeta}$. From Figure \ref{fig:Rmc_sig3}, we find that the threshold is shifted to a larger value, indicating that the rapid spatial variation of $\eta$ leads to an effective diffusivity in the large $|\zeta|$ limit. To find the effective diffusivity arising from the rapid variation in $\eta$, we define an average over small length scales in $r$ denoted by $\left\langle \right\rangle$. Using this average, we write the diffusivity $\eta = \eta_0 + \tilde{\eta}$ where $\eta_0 ( r) = \left\langle \eta ( r) \right\rangle$ is the slowly varying part of the diffusivity while $\tilde{\eta} = \eta(r) - \left\langle \eta( r) \right\rangle$ is the rapidly varying part of the diffusivity. Similarly, we decompose the magnetic field as ${\bf B} = {\bf B}_0 + \tilde{\bf B}$ where ${\bf B}_0 ({\bf r}, t) = \left\langle {\bf B} \right\rangle$ denotes the magnetic field which varies slowly in $r$ while $\tilde{\bf B}$ denotes the magnetic field part which varies rapidly. From the induction equation, on averaging over small $r$, we get the governing equation for the slowly varying magnetic field ${\bf B}_0$ as, 
\begin{align}
    \partial_t {\bf B}_0 = {\bm \nabla} \times \left({\bf u} \times {\bf B}_0 \right) + \eta_0 \nabla^2 {\bf B_0} - \left\langle {\bm \nabla} \times \left(\tilde{\eta} {\bm \nabla} \times \tilde{\bf B} \right) \right\rangle.
    \end{align}
The last term in the above equation is the source of the effective diffusivity of the slowly varying magnetic field ${\bf B}_0$. To find the rapidly oscillating part $\tilde{\bf B}$, we subtract the above equation from the induction equation and consider only the dominant terms leading to
\begin{align}
\eta_0 \nabla^2 \tilde{\bf B} \approx {\bm \nabla} \times \left( \tilde{\eta} {\bm \nabla} \times {\bf B}_0 \right).  \label{eqn:Btilde}
\end{align}
Here $\tilde{\bf B}$ results from the product of rapid variation of diffusivity and the slowly varying current ${\bm \nabla} \times {\bf B}_0$. Taking ${\bf B}_0$ to be the magnetic field for non-varying conductivity ($\zeta = 0$) case we can find $\tilde{\bf B}$ using Eq. \eqref{eqn:Btilde}, in terms of $\tilde{\eta}$ and ${\bf B}_0$. The effective diffusivity denoted by $\eta_e$ is estimated from the spatial average of the ratio $\left\langle {\bm \nabla} \times \left(\tilde{\eta} {\bm \nabla}\times \tilde{\bf B} \right) \right\rangle / \left({\bf \nabla}^2 {\bf B_0} \right)$. The modified dynamo threshold is then given by $\Rm_c \times (1 + \eta_e/\eta_0)$. The dashed line in Fig. \ref{fig:Rmc_sig3} denotes the threshold found above, and it matches well with the numerically found values at large $|\zeta|$ values. 

\section{Large $\Rm$ limit} \label{App:asymptotics}

\begin{figure}[h!]
\centering
 \includegraphics[width=\linewidth]{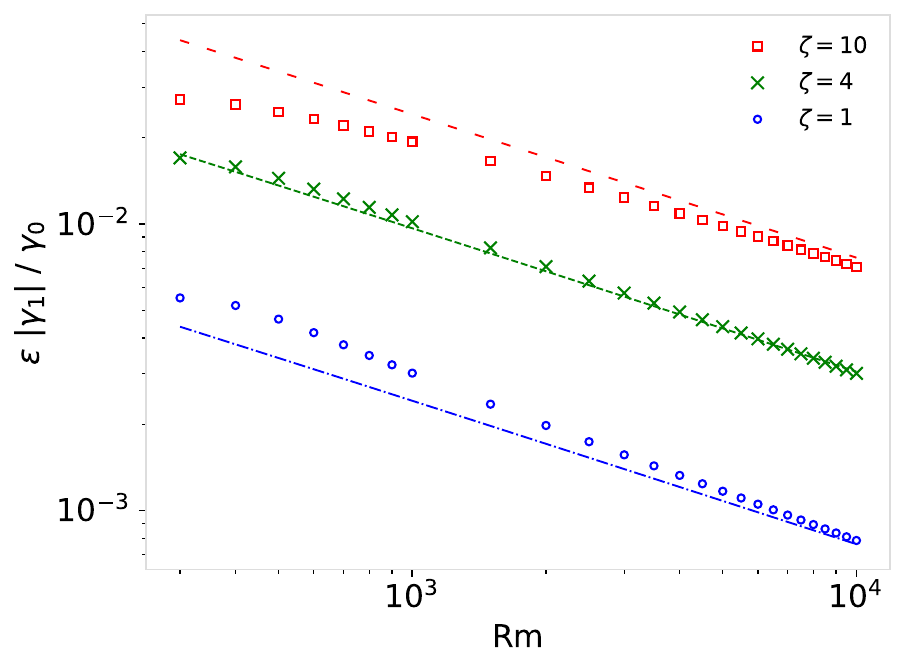}
 \caption{\label{fig:deviation} Figure shows the difference in the growth rate between the cases $\sigma = \frac{3 \sigma_0}{2 + (r/r_0)^{\zeta}}$ and $\sigma = \sigma_0$ as a function of $\Rm$. The lines show the theoretical values found from Eq.\eqref{eq:final_corr} while markers correspond to numerical values. }
\end{figure}

We derive the correction to the growth rate in the limit of large $\Rm$ for the case of the discrete Ponomarenko dynamo. Since the magnetic field gets concentrated at the discontinuity $r = r_0$ as $\Rm \rightarrow \infty$, following \cite{gilbert1988fast}, we expand the quantities in terms of the small parameter $\epsilon$ given by $\eta = \epsilon^2 \eta_0$. The growth rate is expanded as, $\gamma = \epsilon^{-1} \gamma_{-1} + \epsilon^0 \gamma_0 + \epsilon \gamma_1 + \epsilon^2 \gamma_2 + O(\epsilon^3)$. The modes $m$ and $ k$ are taken to scale with $\epsilon$ as $m = m_0 \epsilon^{-1}$ and $k = k_0 \epsilon^{-1}$, respectively. The field is concentrated in a region of width, inversely proportional to the square root of the Reynolds number $\Rm$. Thus, we write $r = r_0 + \epsilon s$. We expand the interior magnetic field in the domain $r \in [0, r_0]$ as,
\begin{align}
    B_r^i (r) = & \epsilon^0 P_0 (s) + \epsilon^1 P_1 (s) + \epsilon^2 P_2 (s) + O(\epsilon^3), \\
    B_\phi^i (r) = & \epsilon^0 Q_0 (s) + \epsilon^1 Q_1 (s) + O(\epsilon^2).
\end{align}
While the exterior magnetic field in the domain $r \in [r_0, \infty)$ are expanded as,
\begin{align}
    B_r^o(r) = & \epsilon^0 R_0(s) + \epsilon^1 R_1(s) + \epsilon^2 R_2(s) +  O(\epsilon^3), \\
    B_\phi^o(r) = & \epsilon^0 S_0(s) + \epsilon^1 S_1(s) + O(\epsilon^2).
\end{align}
The boundary conditions for $P$ and $ Q$ can be obtained using the Eqs.(\ref{eqn:Br_conditions}) and (\ref{eqn:Bp_conditions}),
\begin{align}
    & P_0 \rvert_{r_0} = R_0 \rvert_{r_0}, \; P_1 \rvert_{r_0} = R_1 \rvert_{r_0}, \; P_2 \rvert_{r_0} = R_2 \rvert_{r_0}, \label{eqn:matching_P} \\
    & Q_0 \rvert_{r_0} = S_0 \rvert_{r_0}, \; Q_1 \rvert_{r_0} = S_1 \rvert_{r_0}, \label{eqn:matching_Q} \\
    & P'_0 \rvert_{r_0} = R'_0 \rvert_{r_0}, \; P'_1 \rvert_{r_0} = R'_1 \rvert{r_0}, \; P'_2 \rvert_{r_0} = R'_2 \rvert_{r_0}, \label{eqn:matching_deriv_P} \\
    & \eta_0 Q'_0 \rvert_{r_0} + r_0 \Omega P_1 \rvert_{r_0} = \eta_0 S'_0 \rvert_{r_0}, \; \eta_0 Q'_1 \rvert_{r_0} + r_0 \Omega P_2 \rvert_{r_0} = \eta_0 S'_1 \rvert_{r_0} \label{eqn:matching_deriv_Q}
\end{align}
Finally, substituting the magnetic field expansion into the induction equation and applying the boundary conditions obtained above, we get equations at different orders of $\epsilon$. We reproduce the results already known at $\epsilon^{-1}$ and $\epsilon^0$,  
\begin{align}
    \gamma_{-1} = & 0, P_0 = R_0 = 0, \\
    Q_0 (s) = & c_1 e^{\alpha s}, S_0(s) = c_1 e^{-\alpha s}, \\
    P_1 (s) = & \frac{i m_0 c_1}{r_0^2 \alpha^2} (\alpha s - 1) e^{\alpha s}, \\
    R_1 (s) = & - \frac{i m_0 c_1}{r_0^2 \alpha^2} (\alpha s + 1) e^{- \alpha s}.
\end{align}
where $c_1$ is an arbitrary constant and $\alpha^2 = m_0^2/r_0^2 + k_0^2 + \gamma_0/\eta_0$. The growth rate $\gamma_0$ is found to be,
\begin{align}\label{gamma0}
    \gamma_0 = \frac{\eta_0}{2} |\alpha|^2 - \eta_0 \frac{m_0^2}{r_0^2} \left( 1+ \chi^{-2} \right) + i \frac{\sqrt{3}\eta_0}{2} |\alpha|^2
\end{align}
In the presence of variable conductivity, we get a correction to $\gamma_0$ at the next order. To obtain this correction we solve for $Q_1, S_1$ and $P_2, R_2$ whose governing equations are given by, 
\begin{align}
    Q''_1 - \alpha^2 Q_1 & = \left( \frac{\gamma_1}{\eta_0} - s \frac{\eta'_0}{\eta_0} \frac{\gamma_0}{\eta_0} - s \frac{2 m_0^2}{r^3_0} \right) Q_0 \nonumber \\
    & - \left( \frac{1}{r_0} + \frac{\eta'_0}{\eta_0} \right) Q'_0 , \\
    S''_1 - \alpha^2 S_1 & = \left( \frac{\gamma_1}{\eta_0} - s \frac{2 m_0^2}{r^3_0} \right) S_0 - \frac{S'_0}{r_0}, \\
    P''_2 - \alpha^2 P_2 & = \left( \frac{\gamma_1}{\eta_0} - s \frac{\eta'_0}{\eta_0} \frac{\gamma_0}{\eta_0} - s \frac{2 m_0^2}{r^3_0} \right) P_1 - \frac{P'_1}{r_0} \nonumber \\
    &   + \frac{2 i m_0}{r_0^2} \left( Q_1 - s\frac{2 Q_0 }{r_0} \right) , \\
    R''_2 - \alpha^2 R_2 & = \left( \frac{\gamma_1}{\eta_0} - s \frac{2 m_0^2}{r^3_0} \right) R_1 - \frac{R'_1}{r_0} \nonumber \\
    &     + \frac{2 i m_0}{r_0^2} \left( S_1 - s \frac{2 S_0}{r_0} \right).
    \end{align}
Further, using the matching conditions given in Eqs.\eqref{eqn:matching_P} - \eqref{eqn:matching_deriv_Q}, the correction in the growth rate $\gamma_1$ is found to be,
\begin{align}
\gamma_1 = -\frac{\eta'_0}{\eta_0} \frac{4 \gamma_0 - 3 \alpha^2 \eta_0}{12 \alpha} \label{eq:final_corr}
\end{align}
\sv{This is the same as mentioned in equation \ref{eqn:gamma_corr}}. We then compare the above expression with the numerically found eigenvalues. Figure \ref{fig:deviation} shows the difference in the growth rate between the cases with variable conductivity and without variable conductivity, where Eq.\eqref{eqn:profile_1} gives the conductivity profile. We find a good match between the theoretical solution (lines) obtained from \eqref{eq:final_corr} and the numerical solutions (markers). 
\\

\noindent
It is to be noted that the sign of $\gamma_1$ is purely determined by the sign of $\eta'/\eta$ as the quantity $\frac{4 \gamma_0 - 3 \alpha^2 \eta_0}{12 \alpha}$ is found to have a positive real part for the most unstable mode. Thus, in the large $\Rm$ limit, the modification in the growth rate is determined only by the gradient of the conductivity at the discontinuity $r = r_0$ with a positive $\gamma_1$ when $\eta'(r)$ at $r = r_0$ is negative or equivalently when $\sigma'(r)$ at $r= r_0$ is positive. 


\bibliography{apssamp}

\end{document}